\documentclass[final, authoryear,3p]{elsarticle}

\usepackage{graphicx}

\usepackage{amssymb}

\usepackage{pdfpages}

\journal{Aerosol Science}

\usepackage{siunitx}
\usepackage{booktabs}

\begin{document}

\begin{frontmatter}



\title{Seeded growth of monodisperse and spherical silver nanoparticles}


\author[Metas]{Simon Zihlmann\corref{cor1}\corref{Unibas}}
\ead{zihlmann.simon@gmail.ch}

\author[Metas]{Felix L\"u\"ond}
\ead{Felix.Lueoend@metas.ch}

\author[Metas]{Johanna K. Spiegel}
\ead{Johanna.Spiegel@googlemail.com}

\address[Metas]{Laboratory for Particles and Aerosols, Swiss Federal Institute of Metrology METAS, Lindenweg 50, CH-3003 Bern-Wabern, Switzerland}

\cortext[cor1]{Corresponding author}
\cortext[Unibas]{Now at: Department of Physics, University of Basel, Klingelbergstrasse 82, CH-4056 Basel, Switzerland}

\begin{abstract}
Aiming at spherical and monodisperse silver nanoparticles with diameters up to \SI{100}{nm}, the potential of heterogeneous nucleation of silver particles was explored. Gold seed particles, mainly produced with a spark discharge generator, were carried by nitrogen through a three-zone tube furnace. Silver was evaporated at \SI{1210}{\degreeCelsius} in the first zone of the furnace and particle growth and shaping took place in the subsequent zones, heated to \SI{730}{\degreeCelsius} and \SI{390}{\degreeCelsius} respectively. The generated aerosol was monitored by a scanning mobility particle sizer (SMPS), while parameters, such as furnace temperature, seed particle size and concentration and nitrogen carrier gas flow, were investigated. Off-line atomic force microscopy (AFM) and transmission electron microscopy (TEM) were used to characterize the morphology of the silver nanoparticles in addition to the SMPS scans. Spherical silver nanoparticles with a mobility diameter of more than \SI{115}{nm} and a geometric standard deviation of typically 1.09 or lower at concentrations as large as \SI{5e5}{cm^{-3}} could be produced. The mobility diameter of the monodisperse aerosol could be varied in the range of \SIrange{50}{115}{nm} by changing the furnace temperature or the gold seed particle size. Elemental analysis revealed that the gold from the seed particles formed a homogeneous alloy with the silver ($\leq 3.5$ atomic percent of gold). The growth mechanism could not be identified unambiguously since the obtained silver particles could both originate from heterogeneous nucleation of silver vapour on the seed particles or from coagulation and coalescence of the seed particles with smaller, homogeneously nucleated silver particles. Moreover, the narrow size distribution opens the opportunity to obtain an exclusively singly charged, monodisperse calibration aerosol at sufficient concentrations after and additional mobility selection process.
\end{abstract}

\begin{keyword}
Ag nanoparticle, heterogeneous nucleation, spherical, monodisperse, Au seed particle


\end{keyword}

\end{frontmatter}


\section{Introduction}
\label{sec:Intorduciton}
Metallic nanoparticles are one of the most studied subject within the area of nanoscience. Furthermore, there are already numerous applications in nanotechnology employing metallic nanoparticles, especially silver nanoparticles. Silver nanoparticles are used in many different areas, including metrology (calibration of condensation particle counters (CPCs), CEN TC 264 WG 32), antimicrobial applications \citep{Tran2013}, plasmonic applications \citep{Harra2012} and thermal anchors inside $^3$He/$^4$He dilution fridges \citep{Clark2010}, just to mention a few. Full control over particle size and shape is often demanded by technological applications since the optical, electronic, magnetic and catalytic properties of nanoparticles crucially depend on their size and shape \citep{Byeon2012}. A large surface-to-volume ratio, quantum confinement effects and curvature-induced surface effects are responsible for many novel properties. For example, the size and shape, as well as the dielectric environment of metallic nanoparticles essentially determine the localized surface plasmon resonances of silver nanoparticles \citep{Harra2012}. In metrology, a monodisperse aerosol of known size consisting of spherical particles is required to calibrate differential mobility analyzers (DMAs). Additionally, full traceability in calibration of CPC detection efficiencies requires the use of a monodisperse and singly charged calibration aerosol. Since the particles are classified according to their mobility diameter in aerosol science, spherical particles are a further advantage as it allows to connect the mobility diameter to the real geometric diameter of the particle.

There exist various techniques to generate metallic nanoparticles in general and silver particles in particular. Whereas wet chemical colloidal chemistry and vacuum technologies are associated with drawbacks (e.g. costly equipment) and are tedious to operate, aerosol techniques offer many advantages, such as production of ultra-pure particles (e.g. for noble metal), scalability and an environment friendly and easy production. Commonly, silver nanoparticle aerosols are produced by either liquid flame spray processes \citep{Maekelae2004}, electrospraying of silver colloidal suspensions \citep{Lenggoro2007}, glowing wire generators \citep{Peineke2006}, spark discharge generators \citep{Byeon2008, Tabrizi2009a} or by the evaporation/condensation method of silver in a tube furnace \citep{Scheibel1983}. The latter is one of the simplest methods to generate silver nanoparticles. Thereby, silver is evaporated in a tube furnace and the hot silver vapour, when leaving the furnace, is immediately cooled down, which forces the silver vapour to condense into small primary nanoparticles (a few nanometres in diameter). Due to a high concentration of the primary silver nanoparticles, coagulation, leading to larger and fractal-like nanoparticles, is mostly unavoidable. Heating the silver aerosol in a second furnace transforms the fractal-like particles into compact spheres \citep{SchmidtOtt1988, Shimada1994}. Larger spherical but polydisperse particles (up to $\approx$~\SI{200}{nm}) can be produced by inserting a coagulation volume between the two furnaces \citep{Ku2006}. A monodispere aerosol of spherical silver nanoparticles with a mobility diameter of up to \SI{100}{nm} was recently  produced by \cite{Harra2012} using this technique. In this case however, monodispersity was achieved by the introduction of a DMA in front of the second furnace.

Inspired by the heterogeneous nucleation of water in the atmosphere \citep{Pruppacher1997} and by the SCAR \citep{YliOjanperae2010}, we investigated the potential of heterogenous nucleation of silver. The main difference between homogeneous and heterogeneous nucleation is the absence, respectively, the presence of a nucleation site (e.g. the surface of a seed particle). The energy barrier for the formation of a stable nucleus of the condensed phase is reduced in the case of heterogeneous nucleation and therefore, the nucleation rate is increased \citep{Hinds1982}. The reduced energy barrier for heterogeneous nucleation can also be expressed as a reduced supersaturation ratio, which is needed for an appreciable nucleation rate and subsequent growth of particles. In the case of heterogeneous nucleation, the number of nucleation sites is controlled by the number of seed particles. One can expect heterogeneous nucleation to yield larger particles than homogeneous nucleation, since the available vapour evenly distributes over a smaller number of nucleation sites. Moreover, a controlled nucleation and growth, starting from monodisperse seed particles, can be expected to yield a narrow and controllable size distribution of silver nanoparticles.

Coating of nanoparticles or the production of core-shell nanoparticles is closely related to the heterogeneous generation of nanoparticles. \cite{Boies2011} and \cite{Zdanowicz2013} reported the successful decoration of silica nanoparticles with gold an silver respectively. However, in both cases the seed silica nanoparticles are substantially larger than the metallic nanoparticles on its surface. Furthermore, the metal nanoparticles do not converge into a continuous shell. The gas-phase production of binary alloyed nanoparticles consisting of gallium and gold in varying configuration (core-shell or fully alloyed particles) were reported by \cite{Karlsson2004}.

The potential of heterogenous nucleation of silver in that respect was explored with different seed particle materials and generation methods. The resulting silver nanoparticles were first characterized with a scanning mobility particle sizer (SMPS), consisting of a differential mobility analy7er (DMA) and a condensation particle counter (CPC). The morphology of the nanoparticles was further investigated with off-line atomic force microscopy (AFM) and transmission electron microscopy (TEM).

In this study, we present a novel generation method of silver nanoparticles based on heterogeneous nucleation of silver on gold seed particles, aiming to obtain spherical nanoparticles larger than \SI{100}{nm}. The experimental setup is presented and explained in section \ref{sec:Methods}, whereas section \ref{sec:Results_discussion} contains the results and a discussion.

\section{Materials and methods}
\label{sec:Methods}
\begin{figure}[htbp]
	\centering
	\includegraphics[width=\textwidth]{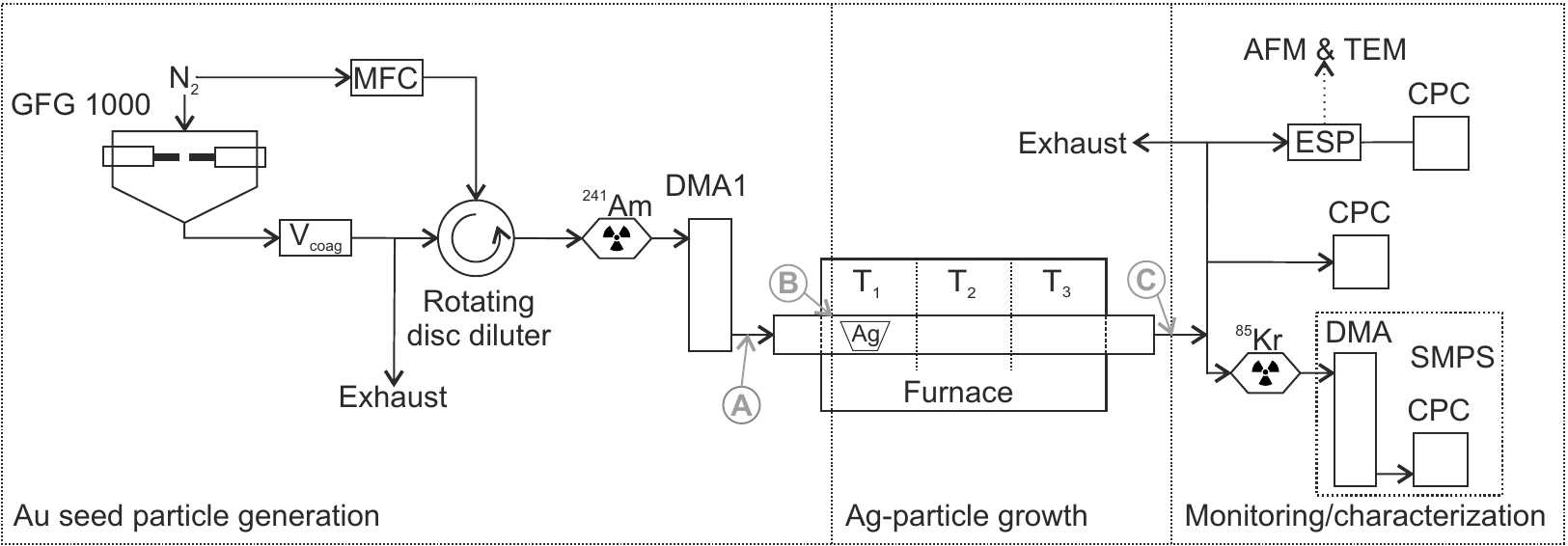}
	\caption{\label{fig:scheme1} Experimental setup, showing the three parts of seed particle generation, growth of the silver particles and their characterization, which are separated from each other by dashed lines. Aerosol particle properties at positions marked with A, B and C in the setup are shown in the SMPS-scans and TEM images in Fig.~\ref{fig:concept}, also marked with A,B and C.
	Abbreviations used: MFC: mass flow controller; V$_{\mathrm{coag}}$: coagulation volume;  DMA: differential mobility analyzer; T: temperature; AFM: atomic force microscopy; TEM: transmission electron microscopy; ESP: electrostatic precipitor; CPC: condensation particle counter; SMPS: scanning mobility particle sizer.}
\end{figure}

The following section explains the generation of heterogeneously nucleated silver particles using a single furnace setup. The generation of the silver particles can be divided into two steps: the formation of appropriate seed particles (e.g. gold seed particles suspended in N$_2$ as a carrier gas) and the subsequent growth of silver particles in a single furnace (see Fig.~\ref{fig:scheme1}). After the generation of the silver particles, the aerosol was analysed and monitored either using an SMPS, a CPC or particles were deposited onto substrates by means of electrostatic precipitation for further off-line analysis.

\subsection{Seed particle generation}
The setup used to generate appropriate gold seed particles is shown in Fig.~\ref{fig:scheme1}. It involved an aerosol generator, some coagulation volumes and dilution steps as well as a DMA to select monodisperse seed particles. At the furnace entrance, the gold aerosol was sintered due to the elevated temperatures before the growth of silver particles took place.

The gold seed particles were produced with a commercially available aerosol generator (GFG 1000, Palas GmbH, Karlsruhe, Germany), originally constructed for the generation of carbon soot particles \citep{Helsper1993}. The working principle of this aerosol generator is based on a spark discharge between two electrodes separated by a small gap (around 2 mm). A \SI{20}{nF} capacitor connected between the two electrodes is charged by a high voltage source with a variable output current. Upon reaching the breakdown voltage of \SI{2}{kV}, the capacitor discharges nearly instantaneously in a spark across the electrode gap, releasing \SI{40}{mJ} of stored energy \citep{Helsper1993}. The spark consists of a high pressure plasma with ionized gas species, reaching local temperatures of up to \SI{20000}{K} \citep{Reimann1997}. Due to the extremely high temperature, electrode material is evaporated in the vicinity of the spark. This vapour is carried away from the electrodes by an inert carrier gas. Rapid cooling of the evaporated material results in high supersaturation ratios, which leads to homogeneous nucleation of the vapour and the formation of small primary particles. Further growth by condensation and coagulation results in highly charged, nanometer-sized agglomerate particles \citep{Schwyn1988}.

In order to generate gold nanoparticles, we followed the modifications proposed by \cite{Messing2009}. The graphite electrodes of the GFG 1000 were replaced by stainless steel rods (\SI{6}{mm} in diameter), functioning as electrode holders for the gold rods (\SI{2}{mm} in diameter, \SI{99.95}{\percent}, GoodFellow, Huntingdon, England). Instead of an argon/air mixture proposed by the manufacturer, nitrogen (\SI{99.999}{\percent}, $\leq$\SI{2}{ppm} O$_2$) was used as an inert carrier gas. Two parameters allow to alter the seed particle size distribution, namely the spark frequency and the carrier gas flow rate. The spark frequency can be varied between 0 and \SI{300}{Hz} and the carrier gas flow from \SIrange{2}{8}{lpm}. The amount of evaporated material scales linearly with the spark frequency \citep{Helsper1993}. For the used settings, see section \ref{subsec:Methods:DefaultSettinggs}.

Downstream of the spark discharge generator, a coagulation volume was installed to increase the particle size of the gold seed particles by further coagulation. Up to a particle mobility diameter of \SI{70}{nm} (unsintered), a coagulation volume of \SI{1.1}{l} was used, above that, a coagulation volume of \SI{2.9}{l} was used. This was necessary as longer residence times in the coagulation volume are needed to achieve particle sizes above \SI{70}{nm}.The particle size distribution of the aerosol was controlled by the spark frequency (amount of available material), carrier gas flow (dilution and coagulation time) and the volume of the coagulation volume (coagulation time). Further downstream of the coagulation volume, a rotating disc diluter (Dilution unit MD 19-1i, Matter Engineering, Wohlen, Switzerland) was installed. This dilution unit with N$_2$ as dilution gas acted as an aerosol source delivering a controlled aerosol flow to the subsequent parts of the setup. This allowed for an independent control of the flow through the GFG 1000 and the flow through the furnace. The excess aerosol flow from the GFG 1000 was fed to an exhaust. The flow delivered by the rotating disc diluter was controlled by a mass flow controller (MFC, 840-L-V1, Sierra Instruments Inc., Monterey, USA).

The DMA1 (M-DMA, 55-340, Grimm, Ainring, Germany) in combination with a $^{241}$Am neutralizer allowed for a size selection of the seed particles. Hereafter, this size is referred to as seed size (e.g. Au80 corresponds to gold seed particles with a mobility diameter of 80 nm prior to sintering). DMA1 was operated at 3 lpm sheath air flow with roughly \SI{1}{lpm} of aerosol flow.  This flow ratio was a trade-off between the monodispersity and concentration of the seed aerosol. To minimize the amount of larger particles having the same electrical mobility due to multiple charges, the seed particles were selected from the descending part of the particle size distribution upstream of DMA1. Nevertheless, a small amount of larger but multiply charged particles can still enter the furnace and act as seed particles.

The size selected aerosol was then passed to the three-zone furnace. Upon entering the furnace, the seed particles were exposed to a high temperature environment and therefore a compaction and sintering process changed the morphology and size of the seed particles, see Fig.~\ref{fig:concept}.

\subsection{Ag-particle growth}
In this part, we explain the experimental setup needed for the growth of the heterogeneously nucleated silver particles (see also Fig.~\ref{fig:scheme1}, central panel).
\\
A three-zone furnace (FRH-3-40/750/1250, Linn High Therm GmbH, Eschenfelden, Germany) represents the main part of this stage. Each of these zones was \SI{250}{mm} long and could be heated independently to temperatures as high as \SI{1250}{\degreeCelsius}. A gas tight ceramic tube (type C610, \SI{60}{\percent} Al$_2$O$_3$) of \SI{1200}{mm} in length and \SI{23}{mm} in diameter (inner) was centred with respect to the middle of the furnace such that on both sides \SI{225}{mm} of the tube was outside of the furnace and therefore not insulated. Inside that tube, a ceramic crucible (Al$_2$O$_3$) filled with a few grams of silver (\SI{99.999}{\percent}, Alfa Aesar, Ward Hill, USA) was positioned in the middle of the first zone of the three-zone furnace, which allowed for a precise control of the silver temperature. A relatively high temperature (\SI{1210}{\degreeCelsius}) in the first zone led to substantial evaporation of silver, which was then available as vapour for the condensation onto the seed particles. The subsequent zones were kept at considerably lower temperatures (\SI{730}{\degreeCelsius} and \SI{390}{\degreeCelsius}) to avoid large temperature gradients, which cause high supersaturation ratios. If the whole furnace is kept at a high temperature, rapid cooling of the silver vapour results in very high supersaturation ratios and hence in homogeneous (i.e. spontaneous) nucleation of a large amount of small silver particles \citep{Scheibel1983}. Such high supersaturation ratios can be avoided by a controlled and smoother temperature gradient from the evaporation temperature (around \SI{1200}{\degreeCelsius}) down to room temperature. The main influences on the saturation profile along the furnace are the three temperatures as well as the nitrogen flow through the furnace.

\subsection{Monitoring and characterization of the aerosol}
\label{subsec:Methods:Characterization}
On-line monitoring and characterization was performed with a TSI SMPS (Neutralizer 3077A, Classifier 3080, DMA 3085 and CPC 3776, lowermost branch in Fig. \ref{fig:scheme1}) connected with a  \SI{1.3}{m} long tube to the outlet of the furnace. Particle size distributions were recorded with a sheath air flow of 3 lpm and an aerosol flow rate of \SI{0.3}{lpm}, ensuring maximal size resolution while keeping the size range of the DMA large enough.
The upscan time of the SMPS was relatively short with \SI{120}{s}. All SMPS scans were corrected for diffusion losses inside the SMPS itself using the built-in correction algorithm in the AIM software from TSI. We did not correct for multiple charges as the charging probabilities in pure nitrogen (as used in this work) differ significantly from the ones in air \citep{Wiedensohler1988}. Since these charging probabilities are implemented in the data inversion algorithm of the AIM, we believe that it is not beneficial to the data to apply an approximative multiple charge correction. To extract the geometric mean diameter $d_P$, the geometric standard deviation $GSD$ and the total concentration $N_0$, a log-normal distribution was fitted to the main peak of the SMPS scans (see Fig. \ref{fig:seeding} for an example of the fits). Even though the log-normal fit underestimates the number of particles in the right part of the peak (tailing), the extracted values do not change significantly compared to the values calculated by the AIM software.
\\
A log-normal distribution can be described as follows:
\begin{equation}
	\label{eq:logn}
	n(d) = \frac{N_0}{\sqrt{2\pi}d}\cdot \exp\left(-\frac{\left(\ln(d) - \mu\right)^2}{2\sigma^2}\right)
\end{equation}
with $N_0$, $\mu$ and $\sigma$ determined from a fit to $n(d)$ data. $N_0$ represents the total concentration, $d_P = e^\mu$ the geometric mean diameter and $GSD = e^\sigma$ the geometric standard deviation, which is a measure of the width of the distribution.

Furthermore, for some experiments, the output of the furnace was monitored with a CPC (3776, TSI) directly (only \SI{9}{cm} of tubing), in order to detect the eventual presence of small homogeneously nucleated silver particles which hardly penetrate the SMPS due to diffusion losses (middle branch in Fig. \ref{fig:scheme1}).

For the off-line characterisation, the particles were sampled with a home-built electrostatic precipitator  (ESP) \citep{Nicolet2012b}. The aerosol flow through the ESP was maintained at \SI{0.3}{lpm} by a CPC (3776, TSI), which also monitored the remaining aerosol concentration during the precipitation (uppermost branch in Fig. \ref{fig:scheme1}). A voltage of +\SI{+3.5}{kV} was applied to collect negatively charged particles. The difference in aerosol concentration in the on and off state was used as a measure of precipitation efficiency, which varied form \SIrange{35}{95}{\percent}, mainly due to different particle sizes. Particles for the AFM analysis were sampled onto freshly cleaved mica whereas particles for TEM analysis were sampled directly onto carbon coated copper grids.

AFM measurements were performed using a modified 'Dimension 3500' metrology AFM from Digital Instruments at METAS. The particle diameter was extracted from the height of the particles, which was measured with respect to the atomically flat mica reference surface, using Scanning Probe Image Processor SPIP \citep{SPIP}. 

TEM images were acquired with a transmission electron microscope (CM12, Philips, Eindhoven, NL) equipped with a digital camera (Morada, Soft Imaging System, M\"unster, Germany) and image analysis software (iTEM) at the Microscopy Imaging Center of the University of Bern as well as at the ZMB at the University of Basel. Scanning transmission electron microscopy (STEM) was used for energy dispersive X-ray (EDX) elemental mapping of the particles. The STEM-EDX was carried out using a FEI Tecnai Osiris with Super-X EDX detector configuration, at 200 kV accelerating voltage and with a \SI{0.9}{nA}, sub-nm probe. The STEM-EDX maps shown are based on the integrated intensity for deconvoluted Au L$_\alpha$ and Ag L$_\alpha$ peaks. High angle annular dark-field (HAADF) images were also recorded.

\subsection{Default settings for the generation of silver particles}
\label{subsec:Methods:DefaultSettinggs}
The following settings were used by default and if nothing else is stated those values were used. The GFG 1000 was operated at a carrier gas flow of \SI{5.1}{slm} and a spark frequency of \SI{300}{Hz}. The coagulation volume downstream of the spark discharge generator was \SI{1.1}{l}. A sheath flow of \SI{3}{lpm} and a voltage of \SI{737}{V} at DMA1 selected gold seed particles with a mobility diameter of \SI{70}{nm}. The flow through the furnace was set to \SI{0.91}{slm}. The three temperatures in the furnace (T$_1$ =\SI{1210}{\degreeCelsius}, T$_2$ = \SI{730}{\degreeCelsius} and T$_3$ = \SI{390}{\degreeCelsius}) were always reached with a ramp rate of \SI{400}{\degreeCelsius\per\hour}.

The standard litre per minutes (slm) are referenced to a temperature of \SI{0}{\degreeCelsius} and to a pressure of \SI{101325}{Pa}.

\section{Results and discussion}
\label{sec:Results_discussion}
This section summarizes the results of the performed experiments. First of all, an example of the heterogeneous nucleation of silver nanoparticles is given (this section). This is followed by a closer look at the growth process, including the influence of the temperature (\ref{subsec:Results:process}),  the seed concentration (\ref{subsec:Results:seed_influence}), the seed size (\ref{subsec:Results:seed_influence}) and the nitrogen carrier gas flow (\ref{subsec:flow_influence_homogeneous}). In the end, homogeneous nucleation of silver (\ref{subsec:Results:Homogeneous}), the charging state (\ref{subsec:Results:Charging_state}) and possible growth mechanisms (\ref{subsec:Results:mechanism}) are discussed.

\begin{figure}[htbp]
	\centering
	\includegraphics{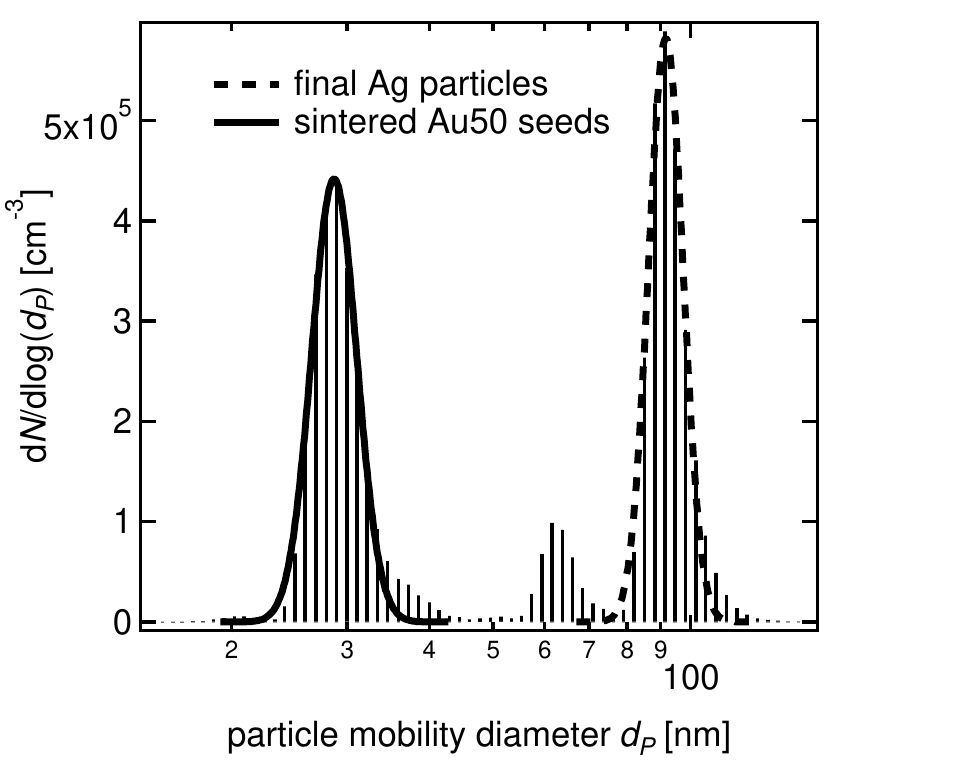} 
	\caption{\label{fig:seeding} Particle size distribution of the sintered Au seeds (50 nm were selected with DMA1, solid line) and the final Ag particles (dashed line). The measured particle size distributions are shown in bars, whereas the log-normal fits to the main peaks in solid and dashed lines. The additional peak around 65 nm is due to doubly charged Ag particles with the size of the main Ag-peak, which appear at smaller diameters in the SMPS scan. The tail towards larger particles of the measured size distribution can partly be reduced if the size distributions are recorded with a longer upscan time ($>$\SI{120}{s}) of the SMPS. However, complete reduction is not accomplished and therefore there are more particles present than suggested by the log-normal fit.}
\end{figure}

The experimental setup described in section \ref{sec:Methods} produces a monodisperse aerosol of spherical silver particles, initiated by gold seed particles (see Fig.~\ref{fig:seeding} for an example with Au50 seeds). Gold seed particles with a mobility diameter of \SI{50}{nm} were selected with DMA1 and passed onto the furnace. Already at \SI{500}{\degreeCelsius}, the gold seed particles completely sintered to compact structures with a size distribution shown in Fig.~\ref{fig:seeding}, fitted with a log-normal distribution shown as solid line. The log-normal fit agrees well with the measured size distribution and allows to extract a geometric mean mobility diameter $d_P$ = \SI{29}{nm} with a $GSD$ = \SI{1.08}{} at a total concentration $N_0$ of \SI{4e4}{cm^{-3}}. Passing the same seed particles to the furnace, but setting the temperatures to the default values (mentioned in \ref{subsec:Methods:DefaultSettinggs}), silver particles with a geometric mean mobility diameter  $d_P$ = \SI{92}{nm}, a $GSD$ = \SI{1.06}{} at a total concentration of \SI{4e4}{cm^{-3}} were obtained (dashed line in Fig. \ref{fig:seeding}). The SMPS did not detect any particles without seed particles entering the furnace (e.g. $U_{\text{DMA1}}$ = \SI{0}{V}), which clearly shows that the gold seed particles are needed for the formation of large silver particles ($\approx$ \SI{100}{nm}) at the default settings. The absence of any particles without seed particles entering the furnace also indicates that there are no unwanted seed particles originating from other sources, e.g. the ceramic tube or crucible.

\subsection{Size and morphology of seeds and final Ag particles}
\label{subsec:Results:process}
\begin{figure}[htbp]
	\centering
	\includegraphics{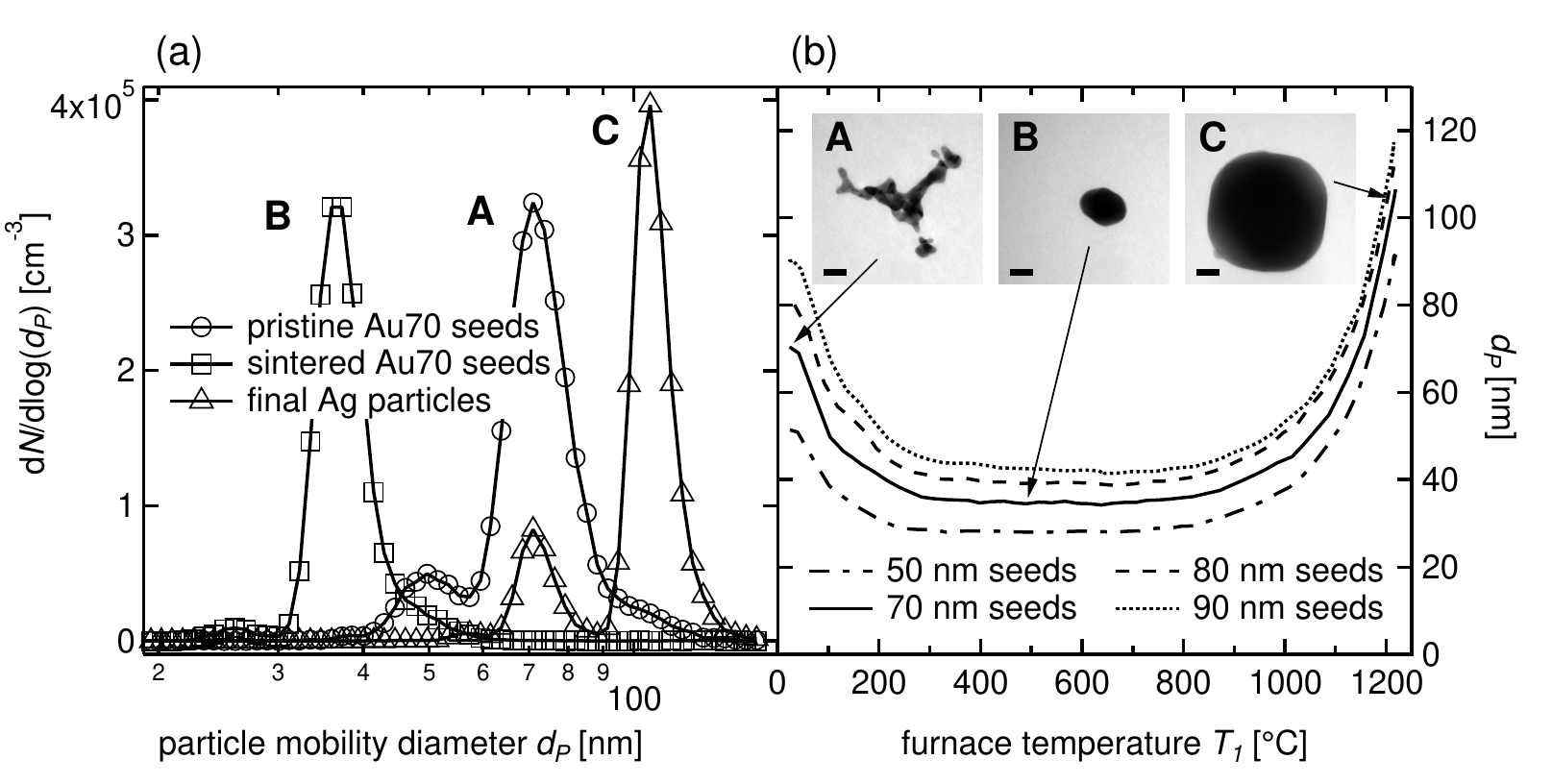}
	\caption{\label{fig:concept} (a) Particle size distributions measured with an SMPS of three different settings, representing pristine Au seeds \textbf{A}, sintered Au seeds \textbf{B} and final silver particles \textbf{C} (see also Fig. \ref{fig:scheme1}). (b) The particle mobility diameter $d_P$ as a function of the furnace temperature $T_1$. The furnace was heated at \SI{400}{\degreeCelsius\per\hour} up to the default values ($T_1$ = \SI{1210}{\degreeCelsius} and $T_2$ ($T_3$) was equal to $T_1$ as long as $T_1 \leq$ \SI{730}{\degreeCelsius} (\SI{390}{\degreeCelsius})). The insets show TEM images of one nanoparticle of each particle size distribution shown in (a). All scale bars are \SI{20}{nm} long and the images are \SI{150}{nm} by \SI{150}{nm} large.}
\end{figure}

Gold seed particles were produced as described in section \ref{sec:Methods}. After the size selection by DMA1 (e.g. \SI{70}{nm} in mobility diameter), a size distribution as shown in Fig.~\ref{fig:concept} (a) \textbf{A} was observed. The particle size distribution selected by DMA1, as depicted in Fig.~\ref{fig:concept} (a), is characterized by a geometric mean mobility diameter of \SI{70.4}{nm} and a $GSD$ of 1.12. Typically, the morphology of nanoparticles produced with a spark discharge generator is far from a compact structure and therefore the mobility diameter is not a precise measure of the geometric dimensions of the particle. An example of such a particle is given in the inset \textbf{A} in Fig.~\ref{fig:concept} (b). The particles are agglomerates of many primary particles forming complex fractal-like and branched structures, which is characteristic for the seed particles leaving DMA1 in our experimental setup (see Fig.~\ref{fig:scheme1}, where the location is indicated by \textbf{A}).

Since it is impossible to monitor the size distribution and morphology of the seed particles at the position indicated with \textbf{B} in Fig.~\ref{fig:scheme1} in our setup, an alternative way of characterizing the size and morphology of our seeds immediately before the growth of the silver particles, had to be chosen. When a seed particle passes through the furnace at the default settings, it first experiences a rapid increase in temperature (first zone) followed by a slow decrease in temperature (second and third zone). The effect of the rapid increase in temperature can be simulated by heating all three zones of the furnace slowly up while monitoring the outlet of the furnace with an SMPS. Fig.~\ref{fig:concept} (b) shows how the particle size $d_P$ evolves with increasing temperature of the furnace. $T_3$ and $T_2$ of the third and second furnace zone were increased at the same rate as $T_1$ until \SI{390}{\degreeCelsius} and \SI{730}{\degreeCelsius}, respectively. As expected, the particle size reduces due to compaction and stabilizes above \SI{300}{\degreeCelsius}. This reduction in particle diameter goes along with a change in morphology, as nicely shown in the inset \textbf{B} in Fig. \ref{fig:concept} (b), which is commonly known as a reshaping and compaction process \citep{Karlsson2005}. The corresponding size distribution of the sintered seed particles at \SI{500}{\degreeCelsius} is shown in Fig. \ref{fig:concept} (a) \textbf{B} with a $d_P$ = \SI{34.5}{nm} and a $GSD$ = \num{1.08}. In comparison to \textbf{A}, the particle mobility diameter has reduced by roughly a factor of two. It is important to note, that at temperatures below \SI{800}{\degreeCelsius} the vapour pressure of silver is too low to result in any growth of particle size nor in particle nucleation. It can be assumed that at the default settings, seed particles undergo a similar change in size and morphology before reaching the region in the furnace where the growth of silver particles takes place (between the first and second zone). The considerably larger dwell time at \SI{500}{\degreeCelsius} in the whole furnace compared to the default settings can be compensated by the much higher temperature of \SI{1210}{\degreeCelsius}.

An increase in $T_1$ from \SI{300}{\degreeCelsius} to \SI{800}{\degreeCelsius} does not change the particle diameter of the gold seed particles. However, internal reconstruction of the crystal structure is known to happen \citep{Karlsson2005}. Above \SI{800}{\degreeCelsius}, a strong increase in $d_P$ is observed. This change in particle mobility diameter is due to a change in $T_1$, since $T_2$ and $T_3$ were constant at \SI{730}{\degreeCelsius} and \SI{390}{\degreeCelsius}. Two main effects, due to an increase in $T_1$, are expected: a larger amount of available silver vapour inside the furnace and a steeper temperature gradient between the first and the second zone in the furnace. The steeper temperature gradient combined with a larger amount of available silver vapour leads to a higher saturation ratio inside the furnace and hence to an enhanced particle growth due to silver deposition onto the gold seed particles. At $T_1$ = \SI{1210}{\degreeCelsius}, a size distribution with $d_P$ = \SI{104}{nm} and a $GSD$ = \num{1.08} is obtained, see also Fig.~\ref{fig:concept} (a) \textbf{C}. Even larger particles are expected for higher $T_1$, which, however, was not tested experimentally. The particle morphology is similar to the sintered seed particles' morphology but the particle size increased considerably (see the insets \textbf{B} and \textbf{C} in Fig.~\ref{fig:concept} (b)). Qualitatively, the temperature dependence of the size and morphology of the particles throughout the furnace is independent on the initial size of the seed particle (see different curves in Fig. \ref{fig:concept} (b)).

In summary, a complex agglomerate gold particle turns into a much larger, compact sphere, mainly consisting of silver, when passed through a furnace with silver vapour present in a well-chosen temperature gradient. At the same time, the temperature profile combined with a possible depletion of silver vapour in the vicinity of seed particles, suppresses homogeneous nucleation of silver vapour, as detectable with an SMPS.

\subsection{Influence of seed particle size and concentration}
\label{subsec:Results:seed_influence}
\begin{figure}[htbp]
	\centering
	\includegraphics{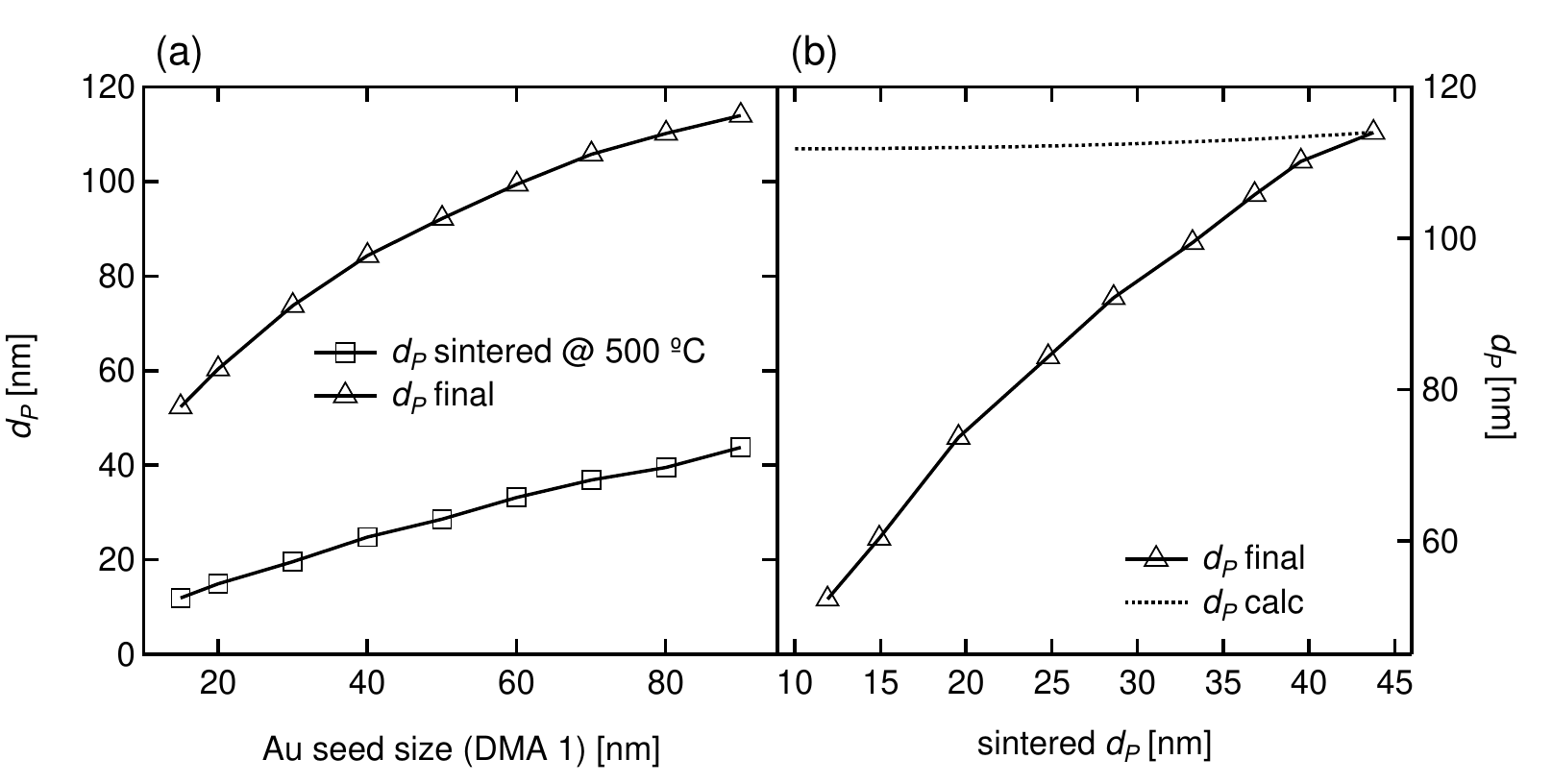}
	\caption{\label{fig:seed_size} Influence of the seed particle diameter on the size of the final silver particles $d_P$. (a) shows the dependence of the sintered seed particle mobility diameter on the initial seed size (given by the DMA1) for the sintering process at \SI{500}{\degreeCelsius} ($\square$, \textbf{B}) and for the final silver particles ($\bigtriangleup$, \textbf{C}, see Fig. \ref{fig:scheme1} and \ref{fig:concept} for details on the capital letters). (b) shows the dependence of the final silver particle mobility diameter on the sintered seed size. The dotted line depicts the final Ag particle size if a constant volume increase would be achieved for all seed sizes (additional assumption: spherical particles).}
\end{figure}

The influence of the size of the gold seed particles on the final silver particles was investigated by changing the voltage of DMA1 and hence the size of the gold seed particles. Since the seed particles undergo a sintering and compaction process, their size after passing the furnace at \SI{500}{\degreeCelsius} was taken as a representative value for their compact size. The particle mobility diameter after sintering follows approximately a linear relationship with nearly a slope of $1/2$ with respect to the selected seed size at DMA1 (see Fig.~\mbox{\ref{fig:seed_size} (a)}). In contrast, the diameter of the final silver particles shows an unexpected strong and sublinear dependence on the seed size (see Fig. \ref{fig:seed_size} (a) and (b)). This dependence is not consistent with a constant volume increase as one would naively assume at first glance. Based on spherical particles, the final particle size was calculated as a function of sintered seed particle diameter, assuming the same volume increase of Ag for all particle sizes by condensation onto the seed particles (example based on the Au90 seeds is shown as dotted line in Fig. \ref{fig:seed_size} (b)). To shed light onto this dependency, detailed computational fluid dynamics simulations including particle growth mechanisms would be needed, which is beyond the scope of the work presented here.
\\
Particle size distribution of the final silver particle of all seed particle sizes together with TEM images are shown in Fig. \ref{fig:seed_size_tem}. All sizes show nearly perfect spherical particles. The tailing towards larger particles of the main peaks is most probably a result of larger seed particles which passed DMA1 with more than one charge. The tailing is enhanced for larger seed sizes which can be explained by the higher probability of multiple charging for larger particles.
\\
A way to reduce the tailing would be to go to smaller seed particles and to increase the furnace temperature, such that more silver is deposited onto the seed particle. However, this approach is limited by the rate of gold evaporation, which is larger for smaller seed particles. In that respect, it might even be a good choice to change to a lower vapour pressure material, e.g. palladium. Using smaller seed particles at a higher furnace temperature would also increase the elemental purity of the final silver particles (see section 3.6).
\\
It has to be noted, that the behaviour in sintering and seeding of the smallest seed particles ($d_P$ = \SI{15}{nm}) deviates significantly from the larger seed particles (see Tab. \ref{tab:seed_size}). First of all, the size distribution of the final silver particles differs from the rest, which was also observed in the larger $GSD$ for the final particle of 1.15 compared to 1.07 for the larger seed particles. Second, a considerable amount ($\approx$ \SI{75}{\percent}) of seed particles was lost inside the furnace. This can most likely be attributed to (partial) evaporation of the smallest gold particles \citep{Nanda2008} inside the furnace. The broader size distribution could also be explained by the fact that the stochastic nature of the nucleation is more pronounced for smaller particles, since the surface is considerably smaller.

\begin{table}[htbp]
	\centering
	\caption{\label{tab:seed_size} Characterisation of the particle size distributions of sintered seed particles and final silver particles. Values of $d_P$, $GSD$ and $N_0$ were obtained by fitting a log-normal distribution to the measured size spectrum (details in section \ref{subsec:Methods:Characterization}). It is important to note that the particle size distribution of the sintered seed particles and the final particle size distribution were not measured at the same time and therefore some variations in $N_0$ can be explained by the not perfectly stable operation of the GFG 1000.}
	\begin{tabular}{l@{\hskip 0.8cm}lll@{\hskip 0.8cm}lll}
		\toprule
			&	\multicolumn{3}{l}{sintered Au seed particles} &	\multicolumn{3}{l}{final Ag particles}\\
		 & $d_{p}$ [nm] & $GSD$ & $N_0$ [cm$^{-3}$] & $d_P$ [nm] & $GSD$ & $N_0$ [cm$^{-3}$] \\
		\midrule
		Au15 & 11.9 & 1.10 & \num{3.4e4} & 52.3  & 1.15 & \num{8.1e3} \\
		Au20 & 14.9 & 1.09 & \num{4.7e4} & 60.3  & 1.10 & \num{4.2e4} \\
		Au30 & 19.5 & 1.08 & \num{4.7e4} & 73.7  & 1.07 & \num{5.3e4} \\
		Au40 & 24.8 & 1.09 & \num{5.5e4} & 84.3  & 1.07 & \num{5.7e4} \\
		Au50 & 28.6 & 1.08 & \num{3.9e4} & 92.1  & 1.06 & \num{3.9e4} \\
		Au60 & 33.2 & 1.08 & \num{4.0e4} & 99.4  & 1.07 & \num{3.9e4} \\
		Au70 & 36.8 & 1.08 & \num{2.7e4} & 105.7 & 1.06 & \num{2.6e4} \\
		Au80 & 39.5 & 1.09 & \num{2.7e4} & 110.2 & 1.08 & \num{2.6e4} \\
		Au90 & 43.8 & 1.10 & \num{2.0e4} & 114.0 & 1.07 & \num{1.7e4} \\
		\bottomrule
	\end{tabular}
\end{table}

\begin{figure}[htbp]
	\centering
	\includegraphics{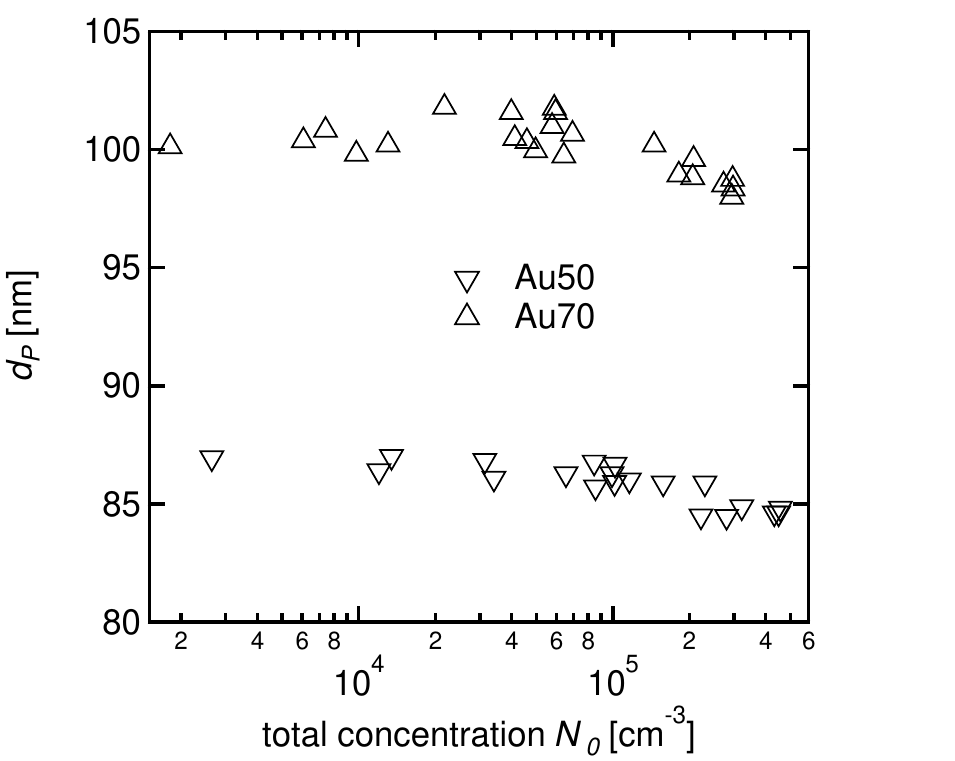} 
	\caption{\label{fig:seed_conc} Influence of the seed particle concentration on the final silver particle mobility diameter $d_P$. Neither 50 nm seeds ($\triangledown$) nor 70 nm seeds ($\vartriangle$) show any significant changes in $d_P$ over two orders of magnitude in total particle concentration $N_0$.}
\end{figure}

\begin{figure}[htbp]
	\centering
	\includegraphics{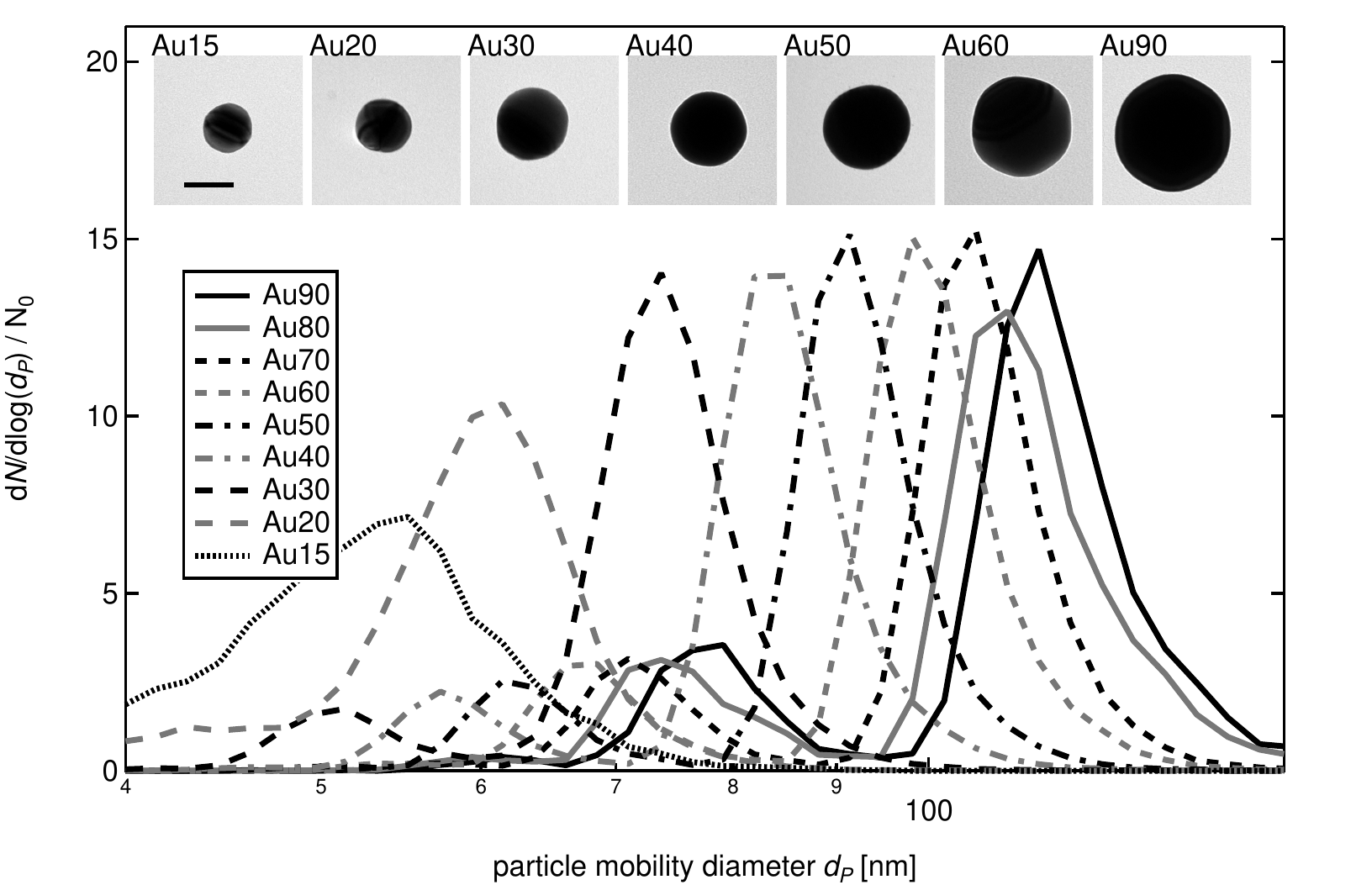} 
	\caption{\label{fig:seed_size_tem} Final silver particle size distributions and exemplary TEM images originating from different seed sizes. All TEM images are \SI{150}{nm} by \SI{150}{nm} in size and the \SI{50}{nm} scale bar of the first images applies to all. Please note that TEM images of the \SI{70}{nm} seed size is shown elsewhere and the \SI{80}{nm} was not sampled due to time limitations. Particles for the TEM images were sampled at default conditions but at a lower flow of \SI{0.65}{slm}, except for the Au15 sample where a flow of \SI{0.91}{slm} was used. The size distributions are normalized with respect to the total concentration $N_0$ and were measured at the default settings without the multiple charge correction. The additional peaks to the left of the main peaks are due to multiply charged particles since the multiple charge correction in the AIM software was deactivated.}
\end{figure}

The concentration of the gold seed particles does not alter the final silver particle size (see Fig.~\ref{fig:seed_conc}). Neither the final silver particles originating from Au50 nor from the Au70 seed particles show any significant change in final particle diameter in the concentration range of of \SIrange{2e3}{5e5}{cm^{-3}}. The fact that the size of the final silver particles is independent on their concentration leads to the conclusion that the size of the final silver particles is not limited by the available amount of silver vapour up to a concentration as high as \SI{5e5}{cm^{-3}}. With our data, we cannot exclude any limitation due to silver vapour depletion in the vicinity of the growing silver particles at concentrations larger than \SI{5e5}{cm^{-3}}. The upper limitation in seed concentration is given by the limited concentration that can be produced with the GFG 1000 and the lower bound is given by the detection limitation of the SMPS.

\subsection{Influence of the nitrogen carrier gas flow}
\label{subsec:flow_influence_homogeneous}
\begin{figure}[htbp]
	\centering
	\includegraphics{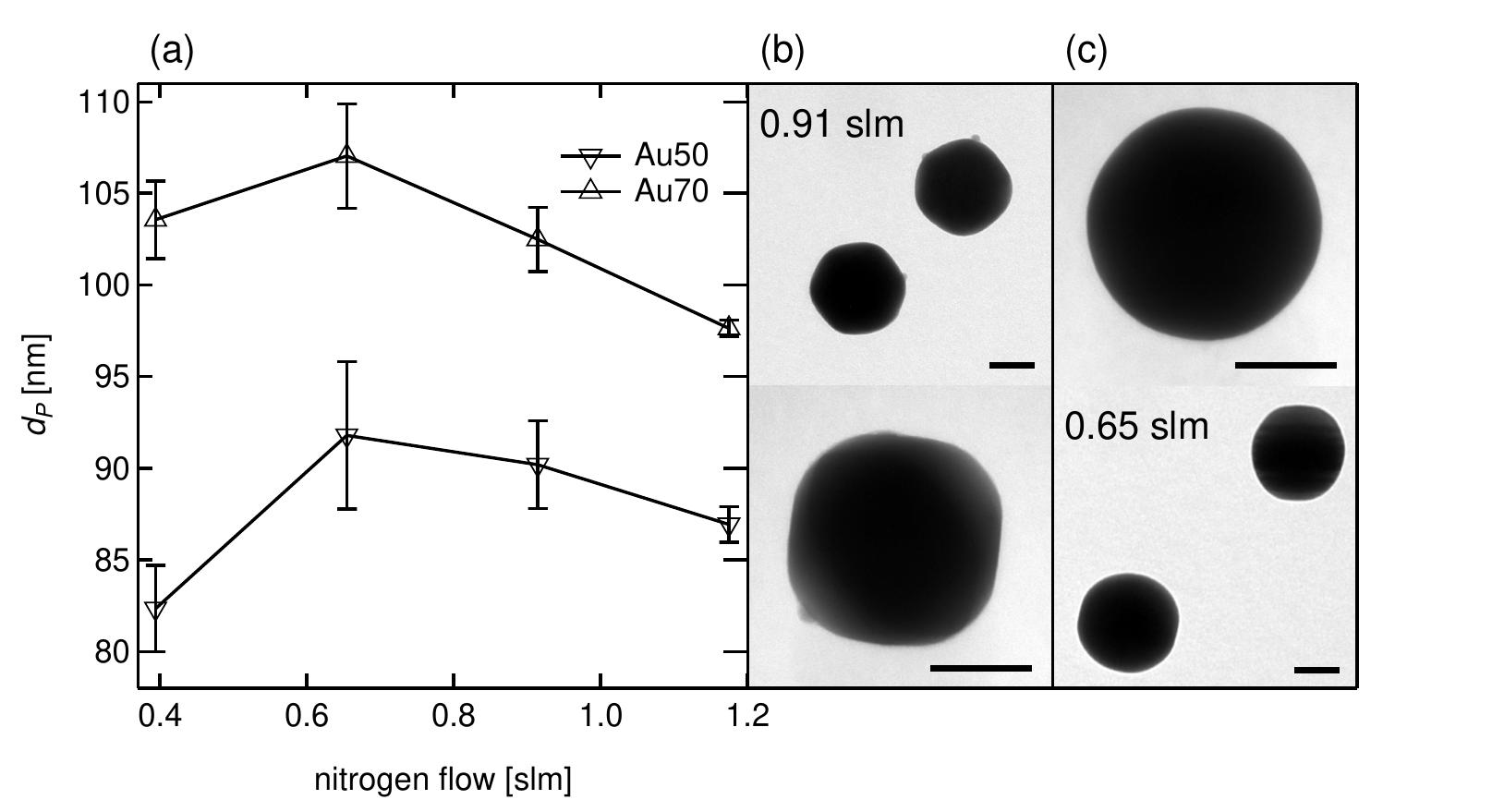} 
	\caption{\label{fig:Q} (a) Influence of the nitrogen carrier gas flow on the the final silver particle size $d_P$. The errorbars are given only by the reproducibility (standard deviation of $d_P$ from six SMPS scans), which is mainly limited by a hysteresis of the particle diameter due to the change of carrier gas flow through the furnace. TEM images of silver nanoparticles produced at the default settings with Au70 seed particles but using a nitrogen flow of \SI{0.91}{slm} (b) and \SI{0.65}{slm} (c). All scale bars are \SI{50}{nm} in length. The roundness of the particles increased with decreasing carrier gas flow and no small particles sticking to the particle surface are observed at \SI{0.65}{slm}.}
\end{figure}

The carrier gas flow through the furnace directly sets the dwell time of the seed particles in the furnace and it also has a great influence on the cooling rate and therefore on the saturation ratio. Basically, a higher carrier gas flow leads to a shorter growth time. However, at the same time a larger supersaturation due to a faster cooling is achieved. For that reason, the influence of the carrier gas flow on the silver particle growth for \SI{50}{nm} and \SI{70}{nm} gold seed particles was investigated in the range of \SIrange{0.39}{1.17}{slm}, see Fig. \ref{fig:Q}. The final silver particle diameter depends only weakly on the carrier gas flow, showing a maximum around \SI{0.65}{slm} (Fig. \ref{fig:Q} (a)). The fact that the particle size does not change significantly when the carrier gas flow is changed by a factor of three, is a hint that the particles grow quickly. Thus, the growth mechanisms have to be much faster than any process related to the carrier gas flow such as drift velocity of the particles in the furnace.

The uncertainties, plotted as errorbars in Fig.~\ref{fig:Q} (a), represent the standard deviation of $d_P$ of six SMPS scans acquired for every gas flow setting. The deviations from one scan to the other is due to a hysteresis in the particle diameter while changing the flow from low to high and to low values again. Since the inner tube surface is by far the largest sink for silver vapour (roughly \num{1e6} larger surface area than the seed particles' surface area), the change in temperature and silver vapour pressure caused by a varied carrier gas flow, results in significant changes in local sinks and sources of silver vapour. Those changes can then easily result in slightly different growth conditions until the equilibrium of evaporation and condensation of silver vapour onto the tube wall is established again.

Interestingly, the nitrogen carrier flow has an effect on the morphology of the final silver particles (Fig. \ref{fig:Q} (b), (c)). At lower flow rates (e.g. \SI{0.65}{slm}) the final silver particles are perfect spheres (see Fig. \ref{fig:Q} (c)) compared to the weakly faceted particle shape at \SI{0.91}{slm}. Furthermore, the randomly distributed tiny particles on the surface of the large silver particles are completely absent for the low flow regime (\SI{0.65}{slm} instead of \SI{0.91}{slm}). These tiny nanoparticles are much smaller than the seed particles. The only possible formation mechanism we could think of would be homogeneous nucleation. Apparently, their presence depends on the nitrogen flow (Fig. \ref{fig:Q} (b), (c)). This can either be explained by the absence of those tiny nanoparticles due to a reduced supersaturation (e.g. not enough for homogeneous nucleation) or that those tiny particles completely coalesced with the large silver particles because of the slower cooling at the reduced flow rate. Therefore, we have a more detailed look at those particles in the next section. 

\subsection{Homogeneous nucleation of silver vapour}
\label{subsec:Results:Homogeneous}
We assume that the small particles visible on the surface of larger particles in the TEM images (Fig. \ref{fig:Q} (b)) and on AFM images (not shown) are formed by additional homogeneous nucleation of silver vapour inside the furnace. Even though AFM images revealed isolated single particles with a mean diameter of \SI{2.5}{nm}, we were not able to locate isolated small particles on the TEM images, using the same sampling procedure. The absence on the TEM images can be attributed to the evaporation of the tiny particle due to electron beam induced heating in the vacuum environment of the microscope. This is further supported by the observation that narrow connections in branched seed particles disappeared during the imaging process. Additionally, small particles sitting on top of large particles were observed (see Fig. \ref{fig:Q} (b)) to disappear after a prolonged time of electron beam irradiation.

However, operating the system at the default settings, none of the tiny homogeneously nucleated silver nanoparticles could be detected by the SMPS, neither with seed particles present, nor without seed particles. The small size of \SI{2.5}{nm} as detected by the AFM might be the reason for this. Interestingly, homogeneously nucleated silver particles were detectable with a CPC (3776, TSI) directly mounted at the outlet of the furnace (only 9 cm tubing to reduce diffusion losses, see also Fig. \ref{fig:scheme1}, middle branch) when the furnace was operated at the default settings without any seed particles. The absence of the seed particles allowed to investigate the formation of those particles with respect to the nitrogen flow and the temperature of the furnace.

The concentration of the homogeneously nucleated particles strongly increased (exponentially) with increasing nitrogen flow through the furnace. Below \SI{0.50}{slm}, the concentration was below \SI{1}{cm^{-3}}. A concentration of roughly \SI{20}{cm^{-3}} was observed at \SI{0.65}{slm} and roughly \num{2e2} particles per \si{cm^{3}} at \SI{0.91}{slm}. It is very important to note that the exact concentrations of those tiny particles is unknown since the counting efficiency of the CPC for the present particles is unknown (D$_{50}$~=~\SI{2.5}{nm}). Flow rate dependent diffusion losses (based on tube penetration efficiencies, \cite{Hinds1982}) do not fully account for the reduced concentration at lower flow rates. This strongly indicates that a higher flow rate itself leads to the formation of more homogeneously nucleated particles. An increased supersaturation in certain regions of the furnace, caused by higher flow rates, is most probably the reason for the observed flow dependence.

Furthermore, a higher $T_1$ lead to a higher concentration of the homogeneously nucleated silver particles. This is obvious since $T_1$ determines the absolute amount of available silver vapour and the temperature gradient along the furnace. Moreover, the concentration of homogeneously nucleated silver particles increases with $T_2$ at fixed $T_1$ and $T_3$ if $T_2$ is higher than \SI{750}{\degreeCelsius}. This dependence on $T_2$ above \SI{750}{\degreeCelsius} suggests that the supersaturated environment corresponding to the origin of formation of those particles is shifted from the interface between zone one and two to the interface of zone two and three of the furnace.
 
In summary, during high flow conditions additional homogeneously nucleated particles with a diameter of a few nanometers were formed. However, in the low flow regime (\SI{0.65}{slm} and lower) the tiny particles were no longer detectable. This information is crucial if the silver aerosol is used in specific applications, such as for calibration purposes in metrology.

\subsection{Charging state of the final Ag-aerosol downstream of a neutralizer}
\label{subsec:Results:Charging_state}
Particularly in metrology, knowledge about the charging state of an aerosol is crucial since it allows traceable calibrations of CPCs with respect to a Faraday cup electrometer (FCAE). Owing to the fact that silver nanoparticles produced with the method presented in section \ref{sec:Methods} fulfil further requirements for a calibration aerosol, such as monodispersity, the question whether the produced Ag aerosol can be singly charged only, was investigated. Detailed information about the methods and the results are found in the supporting materials. In short, we found that the silver nanoparticles, after passing them through a bipolar charger and a DMA with a broad transferfunction centred at the main peak of the size distribution, were not completely singly charged. There were some larger doubly charged particles in the ascending part of the size distribution, that prevent the straight forward usage as a calibration aerosol. However, a well chosen transfer function of a DMA (slightly narrower than the peak and positioned at slightly larger particles than the peak of the size distribution), which is zero in the mobility range of the doubly charged particles allows for the generation of a singly charged, monodisperse aerosol consisting of spherical particles.

\subsection{Elemental analysis}
\label{subsec:Results:elemental}
\begin{figure}[htbp]
	\centering
	\includegraphics[width=3.5cm]{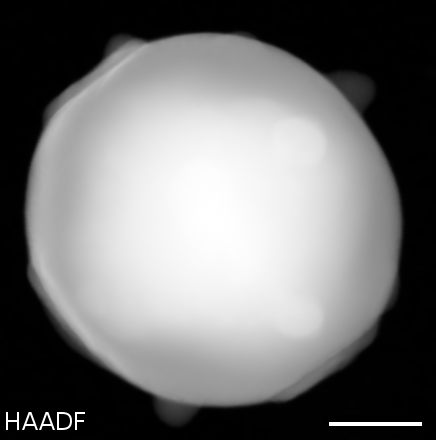}
	\includegraphics[width=3.5cm]{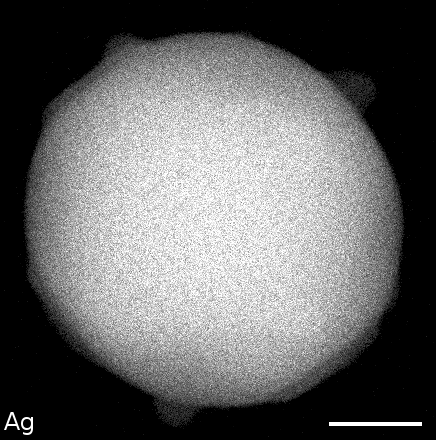}
	\includegraphics[width=3.5cm]{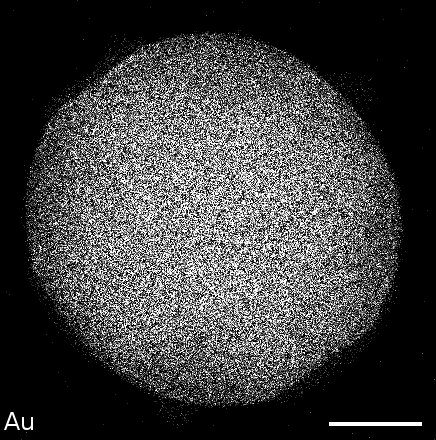}	
	\caption{\label{fig:elemental} A high-angle annular dark-field (HAADF) image and corresponding elemental maps of silver (Ag) and gold (Au) of an Au90 silver nanoparticle are shown. The elemental maps clearly show that the silver and the gold are evenly distributed over the whole nanoparticle and thus form a homogeneous alloy. Furthermore, the HAADF image does not show any contrast difference as one would expect for a core-shell structure. Some additional bumps on the surface of the large nanoparticle are visible, which consist of pure silver. All scale bars are 40 nm.}
\end{figure}

Spatially resolved EDX spectra clearly show that the gold and the silver form a homogeneous alloy throughout the whole nanoparticle, see Fig. \ref{fig:elemental}. An exception are the tiny additional nanoparticles sitting on the surface of larger nanoparticles which only consist of silver. Interestingly, the fraction of gold decreases with decreasing seed size and reaches an undetectable level for the Au30 seeds ($\approx$ \SI{3.5}{atomic percent} and $\approx$ \SI{1.2}{atomic. percent} for Au90 and Au50 respectively). This trend of decreasing gold fraction with decreasing seed size might be attributed to (partial) evaporation of the gold seed during the heterogeneous nucleation of silver. This is further supported by the observation of a reduction of the number concentration for the Au15 sample when the furnace is heated from \SI{500}{\degreeCelsius} to \SI{1210}{\degreeCelsius}, see Tab. \ref{tab:seed_size}. This indicates that the system is truly operated at its limit with gold seeds around \SI{15}{nm} to \SI{30}{nm} and a temperature of \SI{1210}{\degreeCelsius} at nitrogen flow rates around \SI{0.65}{slm} to \SI{0.91}{slm}.
\\
\\
In addition, we found that sampled silver particles on TEM grids are not stable in air for a duration of six months. The shape changed from perfect sphere to a quasi-aggregated structure consisting of several particles. Elemental analysis showed that a considerable amount of sulphur was incorporated into the nanoparticles, indicating the formation of silver sulphide, which is most probably the reason for the change in morphology. It is commonly known that silver tarnishes at ambient conditions and one can imagine that this process is even faster for nanoparticles due to their higher reactivity.

\subsection{Possible growth mechanisms}
\label{subsec:Results:mechanism}
Concerning the growth of the generated silver particles, we could think of basically two growth mechanisms: growth by condensation of silver vapour onto the gold seed particles (heterogeneous nucleation) or growth by coagulation of relatively large gold seed particles with many small homogeneously nucleated silver particles and their subsequent coalescence and sintering into compact spheres.
\\
Apart from the following discussion, we can clearly state that seed particles are necessary for the formation of large, spherical and monodisperse silver nanoparticles and that spontaneous formation of such particles does not take place at the default settings without any seed particles.

The main reason pointing at the coagulation/coalescence growth mechanism is the presence of very small homogeneously nucleated silver particles, which were detectable with a CPC, an AFM and also partly on TEM images (section \ref{subsec:Results:Homogeneous}). Coagulation of particles with different sizes is enhanced compared to a monodisperse aerosol \citep{Hinds1982}. The number of collisions between a seed particle and the homogeneously nucleated silver particles has to be large enough to result in the large final particles. But at the same time, the collisions between the homogeneously nucleated particles themselves are not efficient enough to result in large silver particles. Because the temperature is well above the sintering temperature of silver nanoparticles in the whole furnace \citep{Karlsson2005}, a compact sphere as the final shape of particles formed by coagulation could be expected. The observation of the strong dependence of the final particle size on the seed size also supports the theory of growth by coagulation since coagulation is enhanced for particles with different sizes.

On the other hand, several observations and concepts are not consistent with the growth by coagulation. The independence of the final particle diameter on nitrogen flow is clearly unexpected for a growth model based on coagulation since a lower flow would lead to a longer residence time and therefore to an enhanced number of particle collisions. The fact that the concentration of homogeneously nucleated particles strongly depends on the carrier gas flow rate but the size of the heterogeneously nucleated silver particles not, is a strong evidence that coagulation and coalescence of the small homogeneously nucleated silver particles with seed particles is not the main growth mechanism. This observation therefore points towards heterogeneous nucleation and condensation as the dominating growth process. This is further supported by the fact that heterogeneous nucleation is thermodynamically favoured compared to homogeneous nucleation since it has a lower activation energy. Therefore, heterogeneous nucleation already starts at lower saturation ratios, which are too small for the initiation of homogeneous nucleation. With the temperature gradient along the furnace (default settings) and the nitrogen flow dependent formation of homogeneously nucleated particles, we have strong indications that the saturation ratio is well below traditional evaporation/condensation type systems \citep{Scheibel1983}, especially in the low flow regime. 

To the extend of our knowledge, we cannot rule out any of the above mentioned growth mechanism. It is possible, that both, coagulation and condensation, play a vital role in the growth of the gold seeds to large silver particles.

\section{Conclusion}
\label{sec:Conclusion}
In conclusion, we present a novel and simple method for the generation of spherical silver nanoparticles with a narrow size distribution up to diameters of more than \SI{100}{nm} with a $GSD$ of less than 1.1, initiated by gold seed particles. It is possible to generate heterogeneously nucleated silver particles in concentrations of a few particles up to concentrations of nearly \SI{5e5}{cm^{-3}}. The final silver particle diameter can either be controlled by the temperature of the first zone of the furnace ($T_1$) or by the seed particle diameter. Both offer a flexible way in altering the final particle diameter. Even though the gold from the seed particles forms a homogeneous alloy with the silver, the chemistry of the nanoparticles will still be dominated by the large amount of silver ($> 96$ atomic percent).

The narrow size distribution of heterogeneous nucleation is a great improvement compared to homogeneous nucleation where typical geometric standard deviations are around 1.3 or larger. The novelty of implementing heterogeneous nucleation for silver opens many new possibilities in particle generation, including complex multicomponent nanoparticle synthesis.

The combination of the excellent properties of the silver nanoparticles, generated with the presented setup, make them promising candidates for various applications. Monodisperse, spherical and singly charged particles are needed in metrology (e.g. traceable CPC calibration). The monodispersity and spherical shape could allow for further applications in various other fields, including plasmonic applications for example.

\section{Supplementary material}
\label{sec:SOM}
Detailed information about the charging state of the silver particles downstream of a neutralizer can be found in the supplementary material. The used methods as well as the results are explained in detail. Furthermore, additional seed particle generation methods and seed materials are mentioned.

\section{Acknowledgements}
\label{sec:Acknowledgements}
The authors gratefully thank Ana\"is Nicolet for performing the AFM analysis. Electron microscopy images were acquired on a device supported by the Microscopy Imaging Center of the University of Bern as well as on a device from the Zentrum für Mikroskopie of the University of Basel. The authors want to acknowledge the support from Duncan Alexander at the
Interdisciplinary Center For Electron Microscopy CIME at the EPFL for the elemental analysis. This work was mostly
performed while Simon Zihlmann was doing a civilian service assignment at METAS

\section{Vit\ae}
\label{sec:Vitae}
\subsection{Simon Zihlmann}
\label{subsec:Vitae:Simon}
Simon Zihlmann studied nanoscience at the University of Basel (Switzerland), obtaining his Master of Science in Nanosciences in 2013, under the supervision of Professor Dominik M. Zumb\"uhl, working on hydrogen plasma etching of graphene. During a civilian service assignment and later on during an intern ship, he was working in the Laboratory for Particles and Aerosols at the Swiss Federal Institute of Metrology METAS, where he was working on the growth of heterogeneously nucleated nanoparticles. Since March 2014, he is pursuing a PhD at the University of Basel in the field of quantum transport in graphene.

\subsection{Felix L\"u\"ond}
\label{subsec:Vitae:Felix}
Felix L\"u\"ond studied physics with a focus on solid state physics at ETH Z\"urich and EPF Lausanne, Switzerland. After his diploma in 2005, he worked as a trainee at Kistler Instrumente GmbH for three months. Between 2006 and 2010, he worked at the institute for atmospheric and climate science (IAC) at ETH Zurich, first as a PhD student and subsequently as a postdoctoral fellow. In 2011, he started his position as a junior scientist in the Laboratory for Particles and Aerosols at the Swiss Federal Institute of Metrology METAS. Since June 2013, his quota is shared between the particle laboratory and the laboratory for electrical quantum standards at METAS.
 
\subsection{Johanna K. Spiegel}
\label{subsec:Vitae:Johanna}
Johanna Spiegel studied physics at the University of Augsburg (Germany) and the Norwegian University of Science and Technology in Trondheim (Norway) and received her diploma in 2008 with a diploma thesis investigating the microstructural evolution of snow in the framework of solid state physics. In 2012 she received her PhD from the ETH Zurich for her investigations on fog, focusing on droplet measurement technology and interaction of droplets and Saharan dust. After six months of postdoctoral fellow at ETH Z\"urich, she is now the head of the Laboratory for Particles and Aerosols at the Federal Institute of Metrology METAS.

\bibliographystyle{model5-names}
\bibliography{literature}

\newpage


\section*{Supplementary material}

# Supplementary material
# Seeded growth of monodisperse and spherical silver nanoparticles

Simon Zihlmann[a,*], Felix Lüönd[a], Johanna K. Spiegel[a]

[a]*Laboratory for Particles and Aerosols, Swiss Federal Institute of Metrology METAS, Lindenweg 50, CH-3003 Bern-Wabern, Switzerland*

Additional information about the heterogeneous nucleation of silver nanoparticles is provided. The charging state of the heterogeneously nucleated silver nanoparticles is presented in details, including the methods used. This part shows the possibility and limitations for the silver particles as a calibration aerosol in metrology. In addition, alternative seed particle generation methods and materials are also discussed.

**1. Charging state of the final Ag-particles**

In metrology, the traceable calibration of condensation particle counters (particle concentration standard) is still an issue. Since the response of a condensation particle counter (CPC) and dilution units, for example, are size dependent, a monodispese and singly charged aerosol consisting of spherical particles is needed. Such a perfect aerosol would allow for traceable concentration calibrations with respect to a Faraday cup aerosol electrometer (FCAE) because multiple charge correction involving empirical charging probabilities would become obsolete. Basically, there exist two approaches to generate such a calibration aerosol. First, a singly charged aerosol can be produced by a method developed by Yli-Ojanperä et al. (2010). Second, a particle size distribution being so narrow that the mobility range of the largest multiply charged and the smallest singly charged particles do not intersect, would also be a perfect calibration aerosol. The tunable size, the monodispersity (GSD≈1.07) and the fact that the particles were spherical, made the silver aerosol generated by the method presented in the main article a promising candidate for a calibration aerosol. Due to the high temperatures in the furnace we assume that the silver particles were not only singly charged (Magnusson et al., 1999). Nevertheless, it could be a perfect calibration aerosol if the second criteria would apply to the silver particles. Therefore, we investigated the charging state of the silver particles after a neutraliser using CPC-efficiency measurements (Section 1.1) and tandem differential mobility analyser (TDMA) measurements (Section 1.2).

*1.1. CPC-efficiency measurements*

In order to determine the fraction of multiply charged particles within the size distribution of silver nanoparticles produced with the method presented in the main article (using the default settings), CPC counting efficiency measurements with respect to an FCAE were performed. The CPC counting efficiency was measured as a function of particle size within the main peak of the silver particle size distribution. Assuming that the CPC counting efficiency is not size dependent at this size (e.g. around 100 nm, which is a good assumption since it is well within the plateau region of the CPC), any relative change in CPC counting efficiency could then directly be related to a change of the average charge per particle. This is evident since the FCAE measures the total charge of an aerosol, whereas a CPC measures the number concentration of an aerosol.

---
[*]Corresponding author
  *Email addresses:* `zihlmann.simon@gmail.ch` (Simon Zihlmann), `Felix.Lueoend@metas.ch` (Felix Lüönd),
`Johanna.Spiegel@googlemail.ch` (Johanna K. Spiegel)

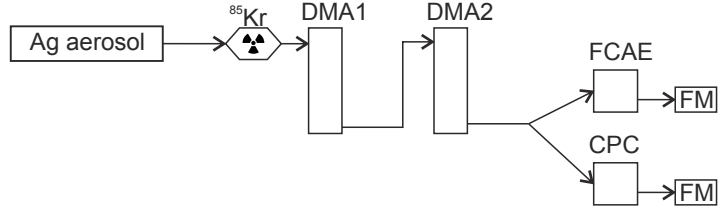

Figure 1: CPC-efficiency measurement. The box labelled with 'Ag aerosol' stands for all the equipment needed to produce silver nanoparticles (see also main text). The aerosol is neutralised with the $^{85}$Kr-neutraliser and then passed through two DMAs. Afterwards the aerosol flow is split and fed to a CPC and a FCAE. The flow through those two instruments was measured with a flow meter (FM).

The setup used to measure the CPC counting efficiency is depicted in Fig. 1 and consists mainly of the aerosol generation unit (Ag aerosol, described in the main article), a bipolar neutraliser, two DMAs, a CPC and an FCAE. The $^{85}$Kr-neutraliser was used to neutralise the aerosol and to provide a known bipolar charge distribution (Wiedensohler et al., 1986; Wiedensohler, 1988). We used DMA1 with an aerosol to sheath air flow of 1.3:3, centred at the maximum of the main peak, to separate the main peak of the heterogeneous nucleated silver particles from potentially available smaller homogeneously nucleated particles (see main article). It is noted here, that the use of DMA1 is not mandatory for this specific experiment. DMA2 with an aerosol to sheath air flow of 1.3:15 was used to select a narrow size fraction of the silver particles with a variable size (see Fig. 2 (a) for three examples of transfer functions). The solid coloured curves depict the narrow transfer functions of particles carrying one charge, whereas the dotted lines depict the transfer functions for doubly charged particles passing DMA2. The CPC counting efficiency was calculated according to ISO 27891:2012, correcting for the flow through each measuring device.

CPC counting efficiency measurements at three different relative positions within the main peak of the silver size distributions are shown in Fig. 1 (b) as blue circles. There exist a clear trend towards higher CPC counting efficiency at larger particles. To exclude a size related bias, the Ag-particle distribution was tuned to a smaller diameter (smaller seeds, red curve) and the CPC counting efficiency was measured again with the same transfer function of DMA2 at 102 nm (red circle). A clear difference exists between the CPC counting efficiency measured in the ascending or in the descending part of the size distribution. This clearly rules out any size related effect and shows that some larger, but doubly charged particles are present in the ascending part of the size distribution. In fact the size distribution of the particles is still large enough that the mobility ranges of singly and doubly charged particles intersect.

*1.2. Tandem DMA*

Further insight into the charging state was obtained using a TDMA setup, shown in Fig 3. A TDMA setup is based on two DMAs, each with a neutraliser upstream that ensures a bipolar charge distribution (Wiedensohler et al., 1986; Wiedensohler, 1988). The aerosol to sheath flow ratio was set to 0.3:3 for DMA1 (3085, TSI) and the second DMA (3081, TSI) was operated with a much narrower transfer function with an aerosol to sheath flow of 0.3:12. These settings ensured that particles passing DMA1 with two charges could be recharged at the second neutraliser such that DMA2 could resolve them as a separate peak in the size spectrum.

Fig. 4 shows a measurement with the TDMA setup. The size distribution upstream of DMA1 is shown in blue, measured with a normal SMPS setup (i.e. without DMA1). Since no multiple charge correction was applied to the SMPS scan, additional peaks due to doubly and triply charged particles are present to the left of the main peak around 113 nm. The red curve indicates the transfer function of DMA1, when operated as a classifier in the TDMA setup. The maximum of the transfer function of DMA1 was set to a particle size lying in the ascending part of the particle size distribution of the heterogeneously nucleated silver nanoparticles. The black bars represent the size distribution measured downstream of the second neutraliser, measured with an SMPS. To the left of the main peak, the fraction of doubly and triply charged particles having the same size are visible as isolated peaks. At larger particle mobility diameters, another isolated peak is

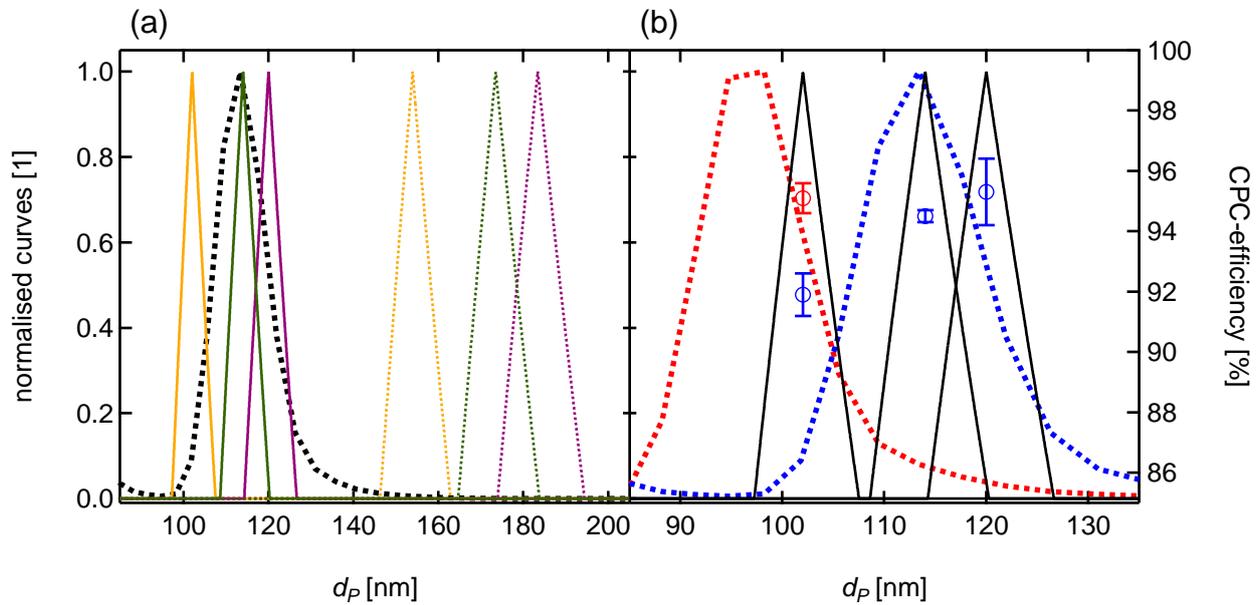

Figure 2: (a) Schematic drawing of the idea of the CPC counting efficiency measurement. The coloured solid lines depict the transfer functions of singly charged particles in the second DMA, whereas the dotted lines correspond to transfer functions of doubly charged particles. This narrow transfer function was scanned stepwise across the main peak (black dotted line) to measure the CPC counting efficiency. For simplicity, all curves are normalized and the left axis applies. (b) Investigation of the charging state of the final Ag-particles. Two Ag-particle size distributions are shown (dashed red and blue curve, both normalized, see axis to the left). The transfer functions of the second DMA are shown as black solid lines and the measured CPC-efficiency thereof is shown as a circle in the corresponding colour (right axis). The errorbars correspond to the uncertainty determined according to ISO 27891:2012.

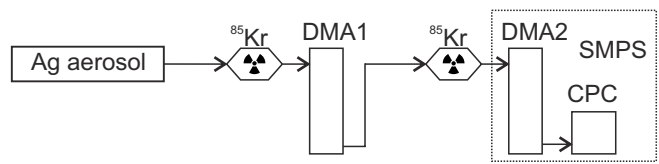

Figure 3: TDMA setup. The box labelled with 'Ag aerosol' stands for all the equipment needed to produce silver nanoparticles (see also main text). Two neutralisers in front of each DMA are essential in this setup because they ensure a bipolar charge distribution before classification with the DMA.

3

visible. This peak corresponds to the fraction of particles carrying two charges that were transmitted by DMA1 together with the singly charged particles. This is a clear evidence that the silver nanoparticles are not only singly charged after a neutraliser even though the size distribution is very narrow. However, we did not observe such a peak due to doubly charged particles when the transfer function of DMA1 was set to the maximum of the size distribution. This opens the door for the generation of a highly monodisperse, singly charged aerosol consisting of spherical silver nanoparticles.

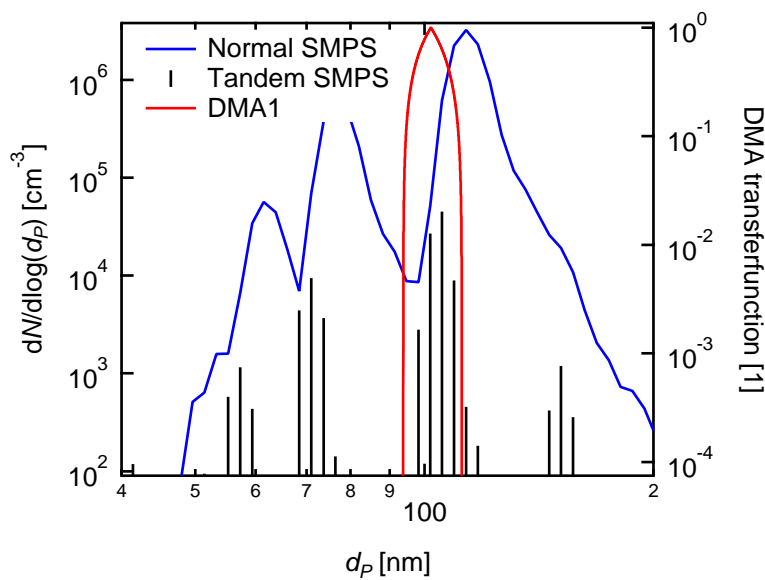

Figure 4: TDMA measurement: the blue curve depicts a normal SMPS scan without multiple charge correction (without the first neutraliser and DMA1 in Fig. 3). The red curve indicates the transfer function of DMA1 in the TDMA setup. The black bars represent the size distribution after neutralization with the second $^{85}$Kr-neutraliser. Clearly, a second peak around 150 nm is observed, which corresponds to particles, which passed the first DMA with two elementary charges. After the second neutrlaiser, they carry only one charge and therefore appear at their true diameter of around 150 nm. The two peaks to the left of the main peak are due to doubly and triply charged particles with a diameter of 102 nm.

In conclusion, the monodisperse silver nanoparticles downstream of a bipolar charger and a mobility selection process (basically removing the pronounced peaks due to doubly and triply charged particles) are not only singly charged. There exists a fraction of doubly charged particles, which have the same mobility as singly charged particles at the lower end the main peak of the silver nanoparticles. This was shown by CPC-efficiency and TDMA measurements.

## 2. Alternative seed particles

In order to get more insight into the formation of heterogeneously nucleated silver nanoparticles, several other seed particle materials and generation methods were investigated in addition to gold.

### 2.1. Generation of alternative seed particles

In general, the generation of seed particles by a spark discharge generator can be applied to any conducting or semiconducting material (Schwyn et al., 1988; Tabrizi et al., 2009). In fact, we also performed some test experiments with copper (99.99+ %, GoodFellow, Huntingdon, England), tungsten (99.97 %, Plansee, Reutte, Austria) and graphite as electrodes in the GFG 1000.

A completely different approach in seed particle generation is the atomisation of a colloid suspension with appropriate seed particles. We performed some experiments with gold colloid suspension from BBInternational, Cardiff, U.K. (30 nm and 40 nm). The gold particle aerosol was produced with an electrospray

aerosol generator (Chen et al., 1995, EAG 3480, TSI, Shoreview, USA). The deployed electrospray generator has been modified during earlier projects. In particular, the radioactive source in the neutraliser has been replaced by a $^{241}$Am source with a considerably longer radioactive half-live. The new pressure control unit allowed for a higher pressure drop across the capillary. The gold colloid suspensions were washed with an ammonium acetate buffer (2 mM, pH 8.1) and concentrated by centrifugation (5000 rpm, 15 min, 3 times) (Böttger et al., 2007). The washing procedure was necessary to reduce the size of the residual particles (originating from anti-coagulant additives in the colloid suspension) to a degree that they are clearly distinguishable from the gold particles in the size spectrum. This assures the uniformity of the seed particles downstream DMA1 in terms of chemical composition. The seed particle concentrations were in the range of few thousand particles per cm$^3$.

2.2. Influence of seed particle material and generation method

Gold nanoparticles originating form electrosprayed gold colloids worked also as seed particles. The only drawback was the limited concentration of a few thousand particles per cm$^3$, which was the main reason for us to switch to the spark discharge generator.

Concerning the spark discharge generator, we also tried copper as a possible material for seed particles. Copper seed particles were produced in the same way as the gold seed particles, namely by replacing the gold rods with copper rods in the GFG 1000. Copper particles showed qualitatively a similar behaviour in sintering and growth as the gold particles below 1000 °C. However, above 1000 °C the particle size decreased rapidly, which we attributed to the evaporation of the seed particles. This is plausible, since the bulk melting temperature of copper is 1083 °C (Lide, 1990-1991) and the vapour pressure of copper is roughly two orders of magnitude higher than the vapour pressure of gold for a given temperature. Furthermore, the sintering and compaction of the copper particles was completed not before 626 °C, which is considerably higher than the 300 °C for gold. Since copper is less noble than gold, it is more reactive and probably possesses more surface contaminations (oxides, carbon contaminations or similar) that hinder the sintering and compaction of agglomerate particles as previously reported for nickel nanoparticles (Seipenbusch et al., 2003).

Additionally, tungsten as the electrode material in the spark discharge generator was also investigated. Two main problems arose during the experiments. Since tungsten has a very high bulk melting temperature, sintering of coagulated particles is expected to be significant only at high temperatures, e.g. a compaction temperature around 1700 °C (Magnusson et al., 2000). This makes it difficult to investigate the seeding efficiency of tungsten particles since a growth in particle size related to silver deposition onto the particles is expected to happen at the same temperatures as sintering and compaction of the tungsten seed particles takes place. Secondly, we observed a broadening of the size distribution combined with a reduction in mobility diameter of the size distribution of the SMPS scans during the sintering of the tungsten nanoparticles (not shown), which we attribute to a breaking apart of the tungsten seed particle into several smaller particles. This could be initiated by the (partly) oxidized tungsten particles itself. Those two problems together with a less stable production of tungsten particles with the GFG 1000, compared to other electrode materials, were reasons enough to focus onto different seed materials, e.g. gold.

We think that palladium might work as well as seed material since its melting point is above gold and more important the vapour pressure is significantly lower than the vapour pressure of gold. Furthermore, spherical palladium particles were successfully produced by a spark discharge generator and a sintering furnace (a compaction temperature around 800 °C was enough) by Messing et al. (2010).

Apart from metallic electrode materials in the GFG 1000, graphite electrodes, as intended by the manufacturer, can also be used to generate seed particles. At first glance, carbon particles might serve as seed particles for the growth of silver nanoparticles since carbon is not melting nor evaporating at temperatures as high as 1210 °C. However, carbon has to be protected against oxidation, which rapidly happens at such high temperatures. The use of nitrogen should prevent oxidation of the carbon particles in our setup. Even though a slight overpressure was present in our setup, we cannot rule out the presence of small quantities of oxygen in our carrier gas which might enter the setup at various positions. We faced similar problems with carbon seed particles as we had with tungsten seed particles. Many small particles were observed while heating up the furnace which made it impossible to investigate the growth of heterogeneously nucleated silver nanoparticles.

## 3. Acknowledgements



\end{document}